\documentclass[tightenlines]{revtex4}
\usepackage{graphicx}% Include figure files
\usepackage{bm}% bold math
\begin{document}

\title{\bf  Ultrafast Auger spectroscopy of quantum well excitons in a strong
  magnetic field}
\author{Tigran V. Shahbazyan}
\affiliation{Department of Physics, Jackson State University, Jackson,
  MS 39217} 

 \begin{abstract}
We study theoretically the ultrafast nonlinear optical 
response of quantum well excitons in a perpendicular magnetic
field. We address the role of many-body correlations originating from the
electron scattering between Landau levels (LL). In the linear optical
response, the processes involving inter-LL transitions are suppressed provided
that the magnetic field is sufficiently strong. 
However, in the {\em nonlinear} response,  the Auger processes
involving inter-LL scattering of {\em two} photoexcited  electrons remain
unsuppressed.  We show that Auger scattering  plays the dominant
role in the coherent exciton dynamics in strong magnetic field. 
We perform numerical calculations for the third-order four-wave-mixing (FWM)
polarization which incorporate the Auger processes nonperturbatively.
We find that inter-LL scattering leads to a strong enhancement and to
oscillations  of the FWM signal at negative time delays. These
oscillations represent quantum beats between optically-inactive
two-exciton states related to each other via Auger processes.
\end{abstract}
%\pacs{PACS numbers: 71.10.Ca, 71.45.-d, 78.20.Bh, 78.47.+p}
\maketitle

\section{introduction}
\label{intro}
It has been established that many-body processes play an important role in the
{\em transient} optical response of semiconductors in the coherent
regime\cite{chemla01,per00,chemla99,axt98}.  The Coulomb correlations between
photoexcited carriers are especially strong in the presence of magnetic field.
By suppressing kinetic energy of electrons and holes in two spatial dimensions
(magnetic confinement), a high magnetic field enhances the relative strength
of interactions between
them\cite{stafford90,stark90,carmel93,rappen91,cundiff96,jiang93,glutsch95}.
In bulk semiconductors, a dominant role of Coulomb correlations in magnetic
field was demonstrated in four-wave mixing (FWM) spectroscopy
experiments\cite{kner97,kner98,kner99}. For example, a huge (several orders of
magnitude) enhancement of the FWM signal was observed as the field exceeded
certain characteristic value. A crossover to strongly-correlated regime
occurs when the magnetic length, $l$, becomes smaller than the excitonic Bohr
radius, $a_B$.

In quantum wells (QW) in perpendicular magnetic field, the energy spectrum is
discrete so one would expect even stronger effect of interactions on the
optical response. The linear absorption spectrum is dominated by a bound
magnetoexciton (MX) state that incorporates electron and hole transitions
between Landau levels (LL) in conduction and valence band, respectively.  In
strong field, such that the cyclotron energy, $\hbar\omega_c$, is much larger
than characteristic interaction energy, $E_0\sim e^2/\kappa l$ (here $\kappa$
is the dielectric constant), the processes involving transitions between
different LL's are suppressed,  and the
lowest MX state is comprised of $n=0$ LL electron-hole ({\em e-h}) pair with
magnetic-field-dependent energy dispersion\cite{butov01}. However, owing to
the {\em e-h} symmetry for any given LL, such MX's do not interact with each
other due to a cancellation of Coulomb matrix elements between electrons and
holes\cite{lerner81,paquet85}.  For this reason, the {\em nonlinear} optical
response of $n=0$ MX's is similar to that of noninteracting two-level
systems\cite{stafford90,stark90,carmel93} unless there is a sufficient 
{\em e-h} asymmetry because of, e. g., differing band offsets or
disorder\cite{bychkov83}.  In the latter case, Coulomb correlations become
important for both pump-probe\cite{chernyak98} or
FWM\cite{yoko99,shahbazyan00,fromer99} spectroscopy; however, in undoped QW's, such an
asymmetry is weak. In weaker magnetic field ($\hbar\omega_c \leq E_0$),
when LL mixing is strong, the coherent optical response 
in QW's was studied in Hartree-Fock approximation (HA) within semiconductor
Bloch equations technique\cite{rappen91,cundiff96}.

Here we study the role of many-body
correlations in coherent optical spectroscopy of QW MX's excited to 
{\em upper} ($n>0$) LL's.  We focus on the case of a sufficiently strong 
magnetic field, $\hbar\omega_c \gg E_0$, so that individual optically-excited
MX, with binding energy $\sim E_0$, is comprised of a single (e. g., $n=1$) LL 
{\em e-h} pair. For such fields, the processes involving inter-LL transitions do
not contribute to linear response even if optical frequency is tuned to excite
interband transitions at upper LL's.  However, as we demonstrate below,
Coulomb correlations between {\em e-h} pairs excited to $n\geq 1$
LL's are significant.  Such correlations originate from Auger processes which
involve inter-LL scattering of {\em two} photoexcited electrons.  For example,
two electrons on $n=1$ LL can scatter to $n=0$ and $n=2$ LL's, as illustrated
in Fig. 1.  Since this is a resonant process (LL's are equidistant), it does
not depend on the LL separation and, therefore, can take place even in a
strong field, $\hbar\omega_c/E_0\gg 1$. The inter-LL Auger processes has been
previously observed in luminescence experiments\cite{potemski91}.

\begin{figure}
\centering
\includegraphics[width=4in]{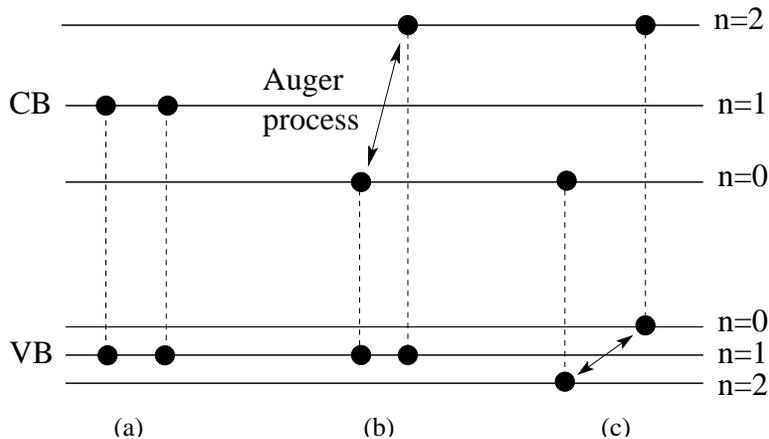}
\caption{\label{fig1} Schematic representation of Auger scattering of
  magnetoexcitons. Two {\em e-h} paires are excited by a pump pulse to
  $n=1$ LL (a) undergo resonant Auger processes involving electrons (b) and
  holes (c).
 }
\end{figure}

We show that the Auger processes play the dominant role in the nonlinear
spectroscopy of QW's in strong magnetic field. Since they involve inter-LL
scattering of charged carriers, the {\em e-h} symmetry no longer holds which
gives rise to interactions between MX's.  It should be emphasized that since
the LL's are discrete, the amplitude of Auger process can be large and, in
fact, is restricted mainly by the inhomogeneous\cite{badalian93,levinson97} or
homogeneous (due to phonons) LL broadening. In fact, in strong field, the
relevant energy scale, $E_0\sim e^2/\kappa l$, is set by interactions, so
that an adequate description of the nonlinear optical response should treat
the Auger processes nonperturbatively. For example, the simple process
described above (Fig. 1) is followed by Auger scattering of the electrons 
{\em back} to $n=1$ LL --- a process completely irrelevant in the
luminescence.  Such {\em multiple} Auger processes, by effectively causing
interaction between MX's, give rise to {\em coherent} signature in the FWM
spectroscopy. We perform numerical calculations of FWM polarization in strong
field by including Auger-scattering {\em exactly}, in all orders of
perturbation theory. We find a strong enhancement as well as oscillations of
the FWM signal for negative time delays. These oscillations are identified as
quantum beats originating from the interference between four-particle
correlated states that are related to each other via Auger processes.

In Section \ref{gen} to we outline the formalism for evaluating the nonlinear
polarization and compute the relevant matrix elements. In Section \ref{auger}
we consider the contribution of the Auger processes into polarization. In
Section \ref{num} we show the results of numerical calculations for the case
of $n=1$ LL. Section \ref{conc} concludes the paper.

\section{Exciton dynamics on arbitrary Landau levels}
\label{gen}
\subsection{General formalism}

We consider a 2D system in strong perpendicular magnetic field with two-band
Hamiltonian $H=H_0+v(r)$, where $H_0$ is free two-band Hamiltonian
and $v(r)$ is the Coulomb potential. In the FWM spectroscopy, the sample is
subjected to the probe and pump laser pulses, separated by time delay $\tau$,  with electric field
intensities ${\cal E}_1(t)$ and  ${\cal E}_2(t)$ and wave vectors ${\bf k}_1$
and ${\bf k}_2$, respectively, and the signal along the direction 
$2{\bf k}-{\bf k}_1$ is measured. In order to obtain the third-order optical
response, we adopt the formalism of Ref. \cite{shahbazyan00} generalized to
the case of arbitrary LL. The third-order FWM optical polarization has
the form 
$P_{FWM}(t)=e^{i(2{\bf k}_2-{\bf k}_1)\cdot {\bf r}}\tilde{P}(t)$
with\cite{shahbazyan00,per00} 
\begin{eqnarray}
\label{pol-fwm}
\tilde{P}(t)=
%&&
i\mu^2 e^{-i\omega_0t}\int_{-\infty}^{t}dt'{\cal E}_1^{\ast}(t')
%\nonumber\\
%&&
\times
\Bigl[\langle 0|T e^{- i H (t - t')}T_{FWM}^{\dag}(t')|0\rangle 
- (t\leftrightarrow t')
\Bigr].
\end{eqnarray}
where $\mu$ is the the interband dipole matrix element, $\omega_0$ is the
laser central frequency (we work in rotating frame), and $|0\rangle$ is the
ground state of $H$. Here 
\begin{equation}
\label{T}
T=\mu\sum_nU_n,
~~~
U_n=\sum_kb_{-kn}a_{kn},
\end{equation}
is the interband transition operator
($a_{kn}$ and $b_{kn}$ are electron and hole annihilation operators,
respectively), while the state $T_{FWM}^{\dag}(t')|0\rangle$ stands for 
\begin{equation}
\label{Tfwm}
T_{FWM}^{\dag}|0\rangle = 
TW^{\dag}|0\rangle - {\cal P}^{\dag}T{\cal P}^{\dag}|0\rangle,
\end{equation}
where the {\em single} {\em e-h} pair state ${\cal P}^{\dag}(t)|0\rangle$ 
and the {\em two} {\em e-h} pair state $W^{\dag}(t)|0\rangle$
satisfy the equations
\begin{equation}
\label{eqP}
i\partial_t{\cal P}^{\dag}(t)|0\rangle 
=H{\cal P}^{\dag}(t) | 0\rangle 
+\mu {\cal E}_2(t)T^{\dag}|0\rangle,
\end{equation}
and
\begin{equation}
\label{eqW}
i\partial_t W^{\dag}(t)| 0 \rangle=H W^{\dag}(t)|0\rangle
+\mu {\cal E}_2(t)T^{\dag}{\cal P}^{\dag}(t)|0\rangle,
\end{equation}
respectively; these equations describe the time-evolution (governed by
the Hamiltonian $H$) of a single-exciton and two-exciton states\cite{shahbazyan00,per00}.

\subsection{Basis}

\subsubsection{Single-exciton states}

The polarization (\ref{pol-fwm}) is determined by
the time-dependence of relevant one and two-pair states. In order to solve
Eqs. (\ref{eqP},\ref{eqW}), we use the standard basis for an electron in $m$th
and a hole in $n$th LL's\cite{kallin84},
\begin{equation}
\label{Psi-exc}
\Psi_{{\bf p}mn}({\bf r}_1, {\bf r}_2)
=N^{-1/2}\sum_ke^{-ikp_xl^2}\psi_{p_y/2+k,m}({\bf r}_1)
\bar{\psi}_{p_y/2-k,n}({\bf r}_2)
=\frac{1}{L}
e^{i{\bf p\cdot R}-iXy/l^2}
\varphi_{mn}({\bf r}+l^2{\bf p} \times {\bf z}),
\end{equation}
where ${\bf p}$ is the
center-of-mass momentum of e-h pair, ${\bf r}={\bf r}_1-{\bf r}_2$, 
${\bf R}=({\bf r}_1+{\bf r}_2)/2$ are the relative and  the 
average coordinates, respectively, $N=L^2/2\pi l^2$ is the LL
degeneracy, and ${\bf z}$ is the unit vector in
the direction of the magnetic field. Here, $\psi_{km}({\bf r}_1)$ and 
$\bar{\psi}_{-kn}({\bf r}_2)=\psi_{kn}^{\ast}({\bf r}_2)$ are the Landau
wave-functions for electron and hole, and $\varphi_{mn}({\bf r})$ is given by 
\begin{equation}
\label{phi-mn}
\varphi_{mn}(z)=
\sqrt{\frac{n!}{m!}}\biggl(\frac{iz}{\sqrt{2}l}\biggr)^{m-n}
L_{n}^{m-n}\biggl(\frac{|z|^2}{2l^2}\biggr)
\frac{e^{-|z|^2/4l^2}}{\sqrt{2\pi l^2}},
~~~
m>n
\end{equation}
and $\varphi_{mn}(z)=\varphi_{nm}(z^{\ast})$ for $m<n$, where
$L_n^{\alpha}(x)$ is the Laguerre polynomial and $z=x+iy$ is the complex 
coordinate. Note also relations 
$\varphi_{mn}^{\ast}(z)=\varphi_{nm}(-z)
=(-1)^{m-n}\varphi_{nm}(z)=(-1)^{m-n}\varphi_{mn}(z^{\ast})$.

In this basis, the single-pair amplitude, 
\begin{eqnarray}
P_{mn}(q,t)=N^{-1/2}\langle {\bf q};mn|{\cal P}^{\dag}|0\rangle, 
\end{eqnarray}
can be easily found from Eq. (\ref{eqP}). The matrix elements of the Coulomb
potential, $v(r)$, have the form
\begin{eqnarray}
\label{V-exc}
V_{mn,m'n'}(p)=
%&&
\langle {\bf p};mn|v|{\bf p};m'n'\rangle
%\nonumber\\
=
%&&
-l^2\int \frac{d{\bf q}}{2\pi}\, v_q\, e^{i{\bf p}\times {\bf q}l^2} 
\varphi_{mm'}(q^{\ast})\varphi_{nn'}(q), 
\end{eqnarray}
where $q=q_x+iq_y$ is the complex
momentum and we used the identity
\begin{equation}
\label{idendity1}
\int d{\bf r} \varphi_{mn}^{\ast}({\bf r})e^{i{\bf q}\cdot{\bf r}} 
\varphi_{m'n'}({\bf r})
=
2\pi l^2 \varphi_{mm'}(q^{\ast})\varphi_{nn'}(q).
\end{equation}
From Eq. (\ref{eqP}), we then obtain 
$P_{mn}({\bf q},t)=\delta_{{\bf q} 0}\delta_{mn}P_{m}(t)$, where $P_{m}(t)$
is the linear polarization, due to the pump, of the $n$th LL, satisfying
\begin{eqnarray}
\label{p-eq}
(i\partial_t -\Omega_m)P_{m}(t)-\sum_n V_{mm,nn}(0)P_{n}(t)=\mu {\cal E}_2(t),
\end{eqnarray}
where $\Omega_{n}=(n+1/2)\omega_c+E_g-\omega_0$ is the detuning ($E_g$ is the
bandgap). 

\subsubsection{Two-exciton states}

Turning to the two-pair states, we introduce the amplitude 
\begin{eqnarray}
W_{mn,m'n'}(p,t)=\langle {\bf p}; mn,m'n'|W^{\dag}|0\rangle, 
\end{eqnarray}
where a complete orthogonal two-exciton basis in the Hilbert subspace with
zero total momentum is constructed from symmetrized product of single-pair states:
\begin{equation}
\label{Psi-twoexc}
\Psi_{{\bf p}mn,m'n'}^{(2)}({\bf r}_1,{\bf r}_2;{\bf r}'_1,{\bf r}'_2)
=\frac{1}{2}\Bigl[
\Psi_{{\bf p}mn}({\bf r}_1,{\bf r}_2)
\Psi_{{\bf -p}m'n'}({\bf r}'_1,{\bf r}'_2)
+\Psi_{{\bf p}mn}({\bf r}'_1,{\bf r}'_2)
\Psi_{{\bf -p}m'n'}({\bf r}_1,{\bf r}_2)
\Bigr],
\end{equation}
The corresponding matrix elements of Coulomb potential can be explicitly
calculated as,
\begin{eqnarray}
\label{V-two}
\langle {\bf p}; mn,m'n'|v|{\bf q};m_1n_1,m'_1n'_1\rangle=
&&
%\frac{1}{2}
\delta_{\bf pq}
\Bigl[\delta_{mm_1}\delta_{nn_1}V_{m'n',m'_1n'_1}(p)
+\delta_{m'm'_1}\delta_{n'n'_1}V_{mn,m_1n_1}(p)
%+(mn)\leftrightarrow (m'n')
\Bigr]
\nonumber\\
&&
+\frac{1}{L^2}\,V_{mn,m'n';m_1n_1,m'_1n'_1}({\bf p},{\bf q}).
\end{eqnarray}
where symmetrization with respect to $(mn)\longleftrightarrow (m'n')$ is
implicit hereafter.
The last term describes MX-MX interaction accompanied by
the momentum exchange
\begin{eqnarray}
\label{V-XX}
V_{mn,m'n';m_1n_1,m'_1n'_1}({\bf p},{\bf q})=
%V_n^{ \alpha \beta,\alpha' \beta'}({\bf p},{\bf q})=
2\pi l^2 v_{|{\bf p} -{\bf q}|}
\Bigl[
e^{-i{\bf p}\times{\bf q}l^2}
\varphi_{mm_1}(\bar{p}-\bar{q})
\varphi_{m'_1m'}(p-q)(-1)^{m'-m'_1}\delta_{nn_1}\delta_{n'n'_1}
\nonumber\\
+e^{i{\bf p}\times{\bf q}l^2}
\varphi_{n_1n}(\bar{p}-\bar{q})
\varphi_{n'n'_1}(p-q)(-1)^{n-n_1}\delta_{mm_1}\delta_{m'm'_1}
\nonumber\\
-\varphi_{mm_1}(\bar{p}-\bar{q})\varphi_{n'n'_1}(p-q)
\delta_{nn_1}\delta_{m'm'_1}
\nonumber\\
-\varphi_{n_1n}(\bar{p}-\bar{q})\varphi_{m'_1m'}(p-q)
\delta_{mm_1}\delta_{n'n'_1}
(-1)^{n-n_1+m'-m'_1}
% \nonumber\\
% +(mn)\leftrightarrow (m'n')
\Bigr],
\end{eqnarray}
where in the derivation we used the relations

\begin{eqnarray}
\label{idendity2}
\int d{\bf r}e^{i{\bf q}\cdot{\bf r}/2}
\varphi_{mn}^{\ast}\Bigl({\bf r}-l^2{\bf q} \times {\bf z}/2\Bigr)
\varphi_{m'n'}\Bigl({\bf r}+l^2{\bf q} \times {\bf z}/2\Bigr)
=
\delta_{mm'}\sqrt{2\pi l^2}\varphi_{nn'}(q),
\nonumber\\
\int d{\bf r}e^{i{\bf q}\cdot{\bf r}/2}
\varphi_{mn}^{\ast}\Bigl({\bf r}+l^2{\bf q} \times {\bf z}/2\Bigr)
\varphi_{m'n'}\Bigl({\bf r}-l^2{\bf q} \times {\bf z}/2\Bigr)
=
\delta_{nn'}\sqrt{2\pi l^2}\varphi_{mm'}(q^{\ast}).
\end{eqnarray}
Equation for two-MX amplitude $W_{mn,m'n'}(q,t)=\langle {\bf p};
mn,m'n'|W^{\dag}|0\rangle$ is 
 obtained by projecting Eq. (\ref{eqW}) onto two-MX basis states,
\begin{eqnarray}
\label{w-eq}
i\partial_t W_{mn,m'n'}(p,t)=
% E_{mnm'n'}^{0}W_{mn,m'n'}(q,t)+
% \sum_{{\bf q}m_1n_1,m'_1n'_1}
% V_{mn,m'n';m_1n_1,m'_1n'_1}({\bf p},{\bf q})
% W_{m_1n_1,m'_1n'_1}(q,t)
\langle {\bf p}; mn,m'n'|HW^{\dag}|0\rangle
%+J_{mn,m'n'}^{0}(p,t),
+{\cal E}_2(t)\langle {\bf p}; mn,m'n'|T^{\dag}{\cal P}^{\dag}|0\rangle,
\end{eqnarray}
where the last (source) term can be found using Eq. (\ref{eqP}),
\begin{equation}
\label{j0}
\langle {\bf p}; mn,m'n'|T^{\dag}{\cal P}^{\dag}|0\rangle
=\frac{\mu}{2}\Bigl[P_{m}(t)+P_{m'}(t)\Bigr]
(N\delta_{{\bf p}0}\delta_{mn}\delta_{m'n'}-\delta_{mn'}\delta_{m'n}),
\end{equation}
where we used the expansion
\begin{eqnarray}
\label{idendity3}
U_m^{\dag}U_n^{\dag}|0\rangle=
N|0;mm,nn\rangle -\sum_{\bf p} |{\bf p};mn,nm\rangle
% \langle {\bf p}; mn,m'n'|T^{\dag}T^{\dag}|0\rangle
% =\mu_{m}\mu_{m'}
% (N\delta_{{\bf p}0}\delta_{mn}\delta_{m'n'}-\delta_{mn'}\delta_{m'n}),
\end{eqnarray}
(the second term comes from the exchange). 

\subsubsection{FWM polarization}

In the following we will be interested in the exciton-exciton interactions
contribution into the FWM polarization. Correspondingly, we separate out 
the interactions-induced contribution by writing
$W^{\dag}|0\rangle=W_{0}^{\dag}|0\rangle+W_{xx}^{\dag}|0\rangle$, where
the first term, corresponding to non-interacting excitons, after being
combined with the second term of (\ref{Tfwm}), gives the Pauli blocking
contribution of non-interacting excitons, and will be included in the
numerical calculations below.
The second term of the above decomposition gives the exciton-exciton
interaction contribution to the polarization,

\begin{equation}
\label{pol-xx}
\tilde{P}^{xx}(t)=
i\mu^2 e^{-i\omega_0t}\int_{-\infty}^{t}dt'{\cal E}_1^{\ast}(t')
\biggl[\phi^{xx}(t,t')-\phi^{xx}(t',t)\biggr],
\end{equation}
with
\begin{equation}
\label{phi-xx}
\phi^{xx}(t,t') = 
\langle 0|T e^{- i H (t - t')}TW_{xx}^{\dag}|0\rangle 
=
\sum_{{\bf p}mnm'n'}
S_{mn,m'n'}({\bf p},t-t')W_{mn,m'n'}(p,t'),
\end{equation}
where the function $S_{mn,m'n'}({\bf p},t)\equiv
\langle 0|T e^{- i Ht}T|{\bf p};mn,m'n'\rangle$
describes the propagation of an {\em e-h} pairs, created by pump and probe
pulses, 
in the FWM direction, and can be expressed via the time-evolution operator
$K(t)$ as 
\begin{equation}
\label{S}
S_{mn,m'n'}({\bf p},t)
% =
% \langle 0|T e^{- i Ht}T|{\bf p};mn,m'n'\rangle
=
\mu^2\sum_{m_1}K_{m_1m}(t)
(N\delta_{{\bf p}0}\delta_{mn}\delta_{m'n'}-\delta_{mn'}\delta_{m'n}),
\end{equation}
with the matrix elements $K_{mn}(t)$ satisfying
\begin{equation}
\label{K}
(i\partial_t -\Omega_{m})K_{mn}(t)
=
\sum_{m_1} V_{mm,m_1m_1}(0)K_{m_1n}(t),
\end{equation}
and $K_{mn}(0)=\delta_{mn}$. Putting all together, we obtain
\begin{equation}
\label{phi-1}
\phi^{xx}(t,t')
=
\mu^2 \sum_{m_1mn}K_{m_1m}(t-t')\chi_{mn}(t'),
\end{equation}
where
\begin{equation}
\label{chi}
\chi_{mn}(t)
=
W_{mm,nn}^{xx}(0,t)-\frac{1}{N}\sum_{\bf p}W_{mn,nm}^{xx}(p,t).
\end{equation}
Substituting the above decomposition
$W_{mn,m'n'}(p,t)=W_{mn,m'n'}^{0}(p,t)+W_{mn,m'n'}^{xx}(p,t)$ into 
Eq.\ (\ref{w-eq}), where the non-interacting term has the form
\begin{equation}
\label{w0}
W_{mn,m'n'}^{0}({\bf p},t)=
\frac{1}{2}P_{m}(t)P_{m'}(t)
(N\delta_{{\bf p}0}\delta_{mn}\delta_{m'n'}-\delta_{mn'}\delta_{m'n}),
\end{equation}
we obtain the following equation for
$W_{mn,m'n'}^{xx}({\bf p},t)$: 
\begin{equation}
\label{wxx-eq}
i\partial_t W_{mn,m'n'}^{xx}(p,t)=
% E_{mnm'n'}^{0}W_{mn,m'n'}(q,t)+
\sum_{{\bf q}m_1n_1,m'_1n'_1}
% V_{mn,m'n';m_1n_1,m'_1n'_1}({\bf p},{\bf q})
% W_{m_1n_1,m'_1n'_1}(q,t)
\langle {\bf p}; mn,m'n'|H|{\bf q}; m_1n_1,m'_1n'_1 \rangle
W_{m_1n_1,m'_1n'_1}^{xx}(q,t)
+J_{mn,m'n'}(p,t),
\end{equation}
where the matrix elements of the Hamiltonian were calculated above and the
source term is given by
\begin{eqnarray}
\label{J}
J_{mn,m'n'}(p,t)=
%&&
\sum_{{\bf q}m_1n_1,m'_1n'_1}
\langle {\bf p}; mn,m'n'|H|{\bf q}; m_1n_1,m'_1n'_1 \rangle
W_{m_1n_1,m'_1n'_1}^{0}(q,t)
\nonumber\\
%&&
-\frac{1}{2}\sum_{m_1}\Bigl[
\Omega_{mm_1}P_{m_1}(t)P_{m'}(t)+\Omega_{m'm_1}P_{m_1}(t)P_{m}(t)
\Bigr]
\nonumber\\
\times
(N\delta_{{\bf p}0}\delta_{mn}\delta_{m'n'}-\delta_{mn'}\delta_{m'n}),
\end{eqnarray}
with $\Omega_{mn}=\Omega_{m}\delta_{mn}-V_{mm,nn}$.
Substitution of Eqs.\ (\ref{V-two}) and (\ref{w0}) into (\ref{J}) yields a
closed-form but rather complicated expression for $J_{mn,m'n'}(p,t)$. In the
following we will need to solve Eqs.\ (\ref{wxx-eq}) only for certain
LL's corresponding to the Auger processes.

%%%%%%%%%%%%%%%%%%%%%%%%%%%%%%%%%%%%%%%%%%%%%%%%%%%%%%%%%%%%

\section{Auger processes}
\label{auger}

We consider the case of strong magnetic field so that $e^2/l\ll \omega_c$. In
this case all {\em single-pair} processes involving inter-LL transitions are
suppressed. For example, neglecting non-diagonal coupling, Eq.\ (\ref{p-eq})
simplifies to
\begin{eqnarray}
\label{p-eq-auger}
(i\partial_t -\Omega_n-V_{nn,nn}(0)+i\Gamma)P_{n}(t)=\mu{\cal E}_2(t),
\end{eqnarray}
and similarly the single-pair evolution operator has only diagonal matrix
elements, $K_{mn}(t)=\delta_{mn}K_n(t)$ with
\begin{eqnarray}
\label{K-auger}
K_n(t)=e^{-i(\Omega_n+V_{nn,nn}(0)-i\Gamma)t},
\end{eqnarray}
where $\Gamma$ is the MX homogeneous width.

In the case when two pairs are excited, inter-LL scattering becomes possible
via Auger processes (see Fig. 1). Let the pump pulse, tuned to $n$th LL, excite two 
{\em e-h} pairs.  Then the 
electrons can Auger-scatter with each other to $n+\alpha$ and
$n-\alpha$ LL's, and holes can Auger-scatter with each other to
$n+\beta$ and $n-\beta$ LL's. Since these are the only {\em resonant}
processes, all other  inter-LL processes are suppressed. 
It is now convenient to introduce new notations which refer to the LL number to
which the {\em e-h} pairs were initially excited,

%%%%%%%%%%%%%%%%%%%%%%%%%%%%%%%%%%%%%%%%%%%%%%%
%
\begin{eqnarray}
\label{energy-auger}
E_{n{\bf p}}^{\alpha \beta \gamma \delta}
\equiv V_{n+\alpha, n+\beta; n+\gamma, n+\delta}(p)
=
-2\pi l^2 \int\frac{d{\bf q}}{(2\pi)^2}
e^{i{\bf p}\times{\bf q}l^2}v_q
\varphi_n^{\alpha \gamma}(\bar{q})
\varphi_n^{\beta\delta}(q),
\end{eqnarray}
and
\begin{eqnarray}
\label{V-XX-auger}
V_n^{ \alpha \beta,\alpha' \beta'}({\bf p},{\bf q})=
2\pi l^2 v_{|{\bf p} -{\bf q}|}
\Bigl[
e^{-i{\bf p}\times{\bf q}l^2}
\varphi_n^{\alpha \alpha'}(\bar{p}-\bar{q})
\varphi_n^{-\alpha' -\alpha}(p-q)(-1)^{\alpha-\alpha'}\delta_{\beta\beta'}
\nonumber\\
+e^{i{\bf p}\times{\bf q}l^2}
\varphi_n^{\beta' \beta}(\bar{p}-\bar{q})
\varphi_n^{-\beta -\beta'}(p-q)(-1)^{\beta-\beta'}\delta_{\alpha\alpha'}
\nonumber\\
-\delta_{\alpha\alpha'}\delta_{\beta\beta'}
[\varphi_n^{\alpha \alpha}(\bar{p}-\bar{q})\varphi_n^{-\beta -\beta}(p-q)
+\varphi_n^{-\alpha -\alpha}(\bar{p}-\bar{q})\varphi_n^{\beta\beta}(p-q)]
\Bigr],
\end{eqnarray}
where 
\begin{eqnarray}
\label{phi}
\varphi_n^{\alpha \beta}(p)
\equiv
\varphi_{n+\alpha,n+\beta}(p)
=\sqrt{\frac{(n+\beta)!}{(n+\alpha)!}}
\Biggl(\frac{ipl}{\sqrt 2}\Biggr)^{\alpha-\beta}
L_{n+\beta}^{\alpha-\beta}(|p|^2l^2/2)
\frac{e^{-|p|^2l^2/4}}{\sqrt {2\pi l^2}},
~~~~
\alpha\geq\beta,
\end{eqnarray}
and $\varphi_n^{\alpha \beta}(p)=\varphi_n^{\beta\alpha}(\bar{p})$ for 
$\alpha < \beta$.

Since all non-Auger inter-LL transitions are suppressed, we restrict
ourselves only with intermediate states related to each other via
Auger processes.  Then the relevant two-MX amplitudes
$w_{n{\bf p}}^{\alpha \beta}(t)
\equiv W_{n+\alpha n+\beta,n-\alpha n-\beta}^{xx}({\bf p},t)$ satisfy the
following system
\begin{eqnarray}
\label{w-eq-auger}
[i\partial_t -2\Omega_n
-E_{n{\bf p}}^{\alpha \beta \alpha \beta}
+E_{n{\bf p}}^{-\alpha -\beta -\alpha-\beta}]
w_{n{\bf p}}^{\alpha \beta}(t)
% \nonumber\\
=\sum_{\alpha' \beta'}\int\frac{d{\bf q}}{(2\pi)^2}
V_n^{ \alpha \beta,\alpha' \beta'}({\bf p},{\bf q})
w_{n{\bf q}}^{\alpha' \beta'}(t)
+m_{n{\bf p}}^{\alpha \beta}(t).
\end{eqnarray}
Note that $w_{n{\bf p}}^{\alpha \beta}(t)=w_{n{\bf p}}^{-\alpha-\beta}(t)$.
The source term $m_{n{\bf p}}^{\alpha \beta}(t)\equiv
J_{n+\alpha, n+\beta;n-\alpha, n-\beta}({\bf p},t)$ can be calculated from 
Eq.\ (\ref{J}) as 
\begin{eqnarray}
\label{source}
m_{n{\bf p}}^{\alpha \beta}(t)
=\Bigl[
v_p\varphi_n^{\alpha \beta}(\bar{p})\varphi_n^{-\beta -\alpha}(p)
(-1)^{\alpha-\beta}(1-\delta_{\alpha\beta})
+E_{n{\bf p}}^{\alpha \beta -\beta -\alpha}
(-1)^{\alpha+\beta}(1-\delta_{\alpha+\beta})
\Bigr]
\nonumber\\
\times
\frac{1}{2}\Bigl[P_{n+\alpha}(t)P_{n-\alpha}(t)
+P_{n+\beta}(t)P_{n-\beta}(t)\Bigr],
\end{eqnarray}
where we used $W_{n+\alpha, n+\beta;n-\alpha, n-\beta}^{0}({\bf p},t)
=\frac{1}{2}P_{n+\alpha}P_{n-\alpha}(N\delta_{\bf p}
\delta_{\alpha\beta}-\delta_{\alpha+\beta})$.
In the absence of Auger scattering ($\alpha=\beta=0$), 
we have $w_{n0}^{00}(t)=0$, i.e. 
interactions do not contribute to the FWM signal (in the ideal 2D case). 

Using  that $w_{n{\bf p}}^{\alpha \beta}(t)$ is related to its Fourier
transform as  
$w_{n{\bf p}}^{\alpha \beta}(t)=-N^{-1}\sum_{\bf q} e^{i{\bf q \times p}l^2}
w_{n{\bf q}}^{\alpha -\beta}(t)$, we obtain from Eq. (\ref{chi}) that
\begin{eqnarray}
\label{chi-auger}
\chi_n^{\alpha}(t)\equiv \chi_{n+\alpha n-\alpha}(t)=2w_{n0}^{\alpha
  \alpha}(t). 
\end{eqnarray}
Using Eqs (\ref{chi-auger}) and (\ref{K-auger}), the FWM polarization can
be  easily expressed in terms of amplitudes $w_{n0}^{\alpha \alpha}(t)$. In
the following, we consider the case when the spectral width of the pulse is
smaller than the LL separation. In this case, we have 
$P_{n\pm\alpha}=\delta_{\alpha 0}P_n(t)$, and the FWM polarization takes
the form (restoring Pauli blocking contribution)
\begin{equation}
\label{FWM-gen}
\tilde{P}(t)=\tilde{P}^{xx}(t)+\tilde{P}^{PB}(t),
\end{equation}
with
\begin{equation}
\label{Pxx-auger-final}
\tilde{\cal P}^{xx}(t,\tau)=i\mu^2\int _{-\infty}^{t}
dt^{\prime}
e^{-\Gamma|t-t^{\prime}|-i\omega_0(t+t^{\prime})}{\cal E}_{1}(t')
%\sum_{\alpha}
\Bigl[
e^{-i(\Omega_n+E_{n0}^{00})(t-t^{\prime})}
\chi_n^{0}(t^{\prime})
-
e^{i(\Omega_n+E_{n0}^{00})(t-t^{\prime})}
\chi_n^{0}(t)
\Bigr],
\end{equation}
and
\begin{eqnarray}
\label{PB}
\tilde{\cal P}^{PB}(t,\tau)=
%&&
i\mu^2\int _{-\infty}^{t}
dt^{\prime}{\cal E}_{1}(t')
e^{-i\omega_0(t+t^{\prime})}
\Bigl[e^{-\Gamma|t-t^{\prime}|-i(\Omega_n+E_{n0}^{00})(t-t^{\prime})}
P_n^2(t^{\prime})-P_n(t)P_n(t^{\prime})
\Bigr]
\nonumber\\
%&&
+\mu^3 P_n(t)
\int_{-\infty}^{t}dt^{\prime}{\cal E}_{2}(t')
\int_{-\infty}^{t^{\prime}}dt''
{\cal E}_{1}(t'')
e^{-i\omega_0(t+t'')+i(\Omega_n+E_{n0}^{00})(t'-t'')
-\Gamma|t'-t''|},
\end{eqnarray}
where $E_{n0}^{00}\equiv V_{nn,nn}(0)$ and
\begin{eqnarray}
\label{P-final}
P_n(t)=
-i\mu \int_{-\infty}^{t}dt' {\cal E}_2(t')
e^{-i(\Omega_n+E_{n0})(t-t')-\Gamma|t-t'|}
\end{eqnarray}
is the pump-induced polarization on $n$th LL.

\section{Discussion and numerical results}
\label{num}

We consider the case when the central laser
frequency is tuned to interband transition between $n=1$ LL's. In this case,
two exciton excited by the pump to $n=1$ LL, can Auger-scatter only
to $n=0$ and $n=2$ LL's, i.e., $\alpha,\beta=0,\pm 1$ (see Fig. 1).
System (\ref{w-eq-auger}) then takes the form:
\begin{eqnarray}
\label{eq-n=1}
&&
(i\partial_t -2\Omega_1-2E_{1p}^{00})w_{1p}^{00}(t)
-\int\frac{d{\bf q}}{(2\pi)^2}V_1^{00}({\bf p},{\bf q})w_{1q}^{00}(t)
-2\int\frac{d{\bf q}}{(2\pi)^2}
\tilde{V}_1^{01}({\bf p},{\bf q})w_{1q}^{01}(t)
=0,
\nonumber\\
&&
(i\partial_t -2\Omega_1-E_{1p}^{11}-E_{1p}^{-1-1})w_{1p}^{11}(t)
-\int\frac{d{\bf q}}{(2\pi)^2}V_1^{11}({\bf p},{\bf q})w_{1q}^{11}(t)
-\int\frac{d{\bf q}}{(2\pi)^2}
\tilde{V}_1^{01}({\bf p},{\bf q})w_{1q}^{01}(t)
=0,
\nonumber\\
&&
(i\partial_t -2\Omega_1-2E_{1p}^{1-1})w_{1p}^{1-1}(t)
-\int\frac{d{\bf q}}{(2\pi)^2}V_1^{1-1}({\bf p},{\bf q})w_{1q}^{1-1}(t)
-\int\frac{d{\bf q}}{(2\pi)^2}
\tilde{V}_1^{01}({\bf p},{\bf q})w_{1q}^{01}(t)
=0,
\nonumber\\
&&
(i\partial_t -2\Omega_1-E_{1p}^{01}-E_{1p}^{0-1})w_{1p}^{01}(t)
-\int\frac{d{\bf q}}{(2\pi)^2}V_1^{01}({\bf p},{\bf q})w_{1q}^{01}(t)
-\int\frac{d{\bf q}}{(2\pi)^2}
\tilde{V}_1^{01}({\bf p},{\bf q})[w_{1q}^{00}(t)
\nonumber\\
&&
\mbox{\hspace{3.5in}}
+w_{1q}^{11}(t)+w_{1q}^{1-1}(t)]
=m_{1p}^{01}(t),
%\nonumber
\end{eqnarray}
where the source term and inter-pair Coulomb matrix elements are given by
\begin{eqnarray}
%\label{eq-n=1}
&&
m_{1p}^{01}(t)=\Biggl[\frac{v_p}{2\pi l^2}e^{-p^2l^2/2}
\Bigl(\frac{pl}{2}\Bigr)^2 L_1^1(p^2l^2/2)-\tilde{E}_{1p}^{01}\Biggr]
\frac{P_1^2(t)}{2},
\nonumber\\
&&
V_1^{00}({\bf p},{\bf q})=-4\sin^2({\bf p\times q}l^2/2)
e^{-|{\bf p-q}|^2l^2/2}v_{|{\bf p-q}|}\bigl[L_1(|{\bf p-q}|^2l^2/2)\bigr]^2,
\nonumber\\
&&
V_1^{11}({\bf p},{\bf q})=-4\sin^2({\bf p\times q}l^2/2)
e^{-|{\bf p-q}|^2l^2/2}v_{|{\bf p-q}|}L_2(|{\bf p-q}|^2l^2/2),
\nonumber\\
&&
V_1^{1-1}({\bf p},{\bf q})=
-e^{-|{\bf p-q}|^2l^2/2}v_{|{\bf p-q}|}
\biggl[ 4\sin^2({\bf p\times q}l^2/2)L_2(|{\bf p-q}|^2l^2/2)
\nonumber\\
&&\mbox{\hspace{2.5in}}
+\Bigl[1-L_2(|{\bf p-q}|^2l^2/2)\Bigr]^2
\biggr],
\nonumber\\
&&
V_1^{01}({\bf p},{\bf q})=
-e^{-|{\bf p-q}|^2l^2/2}v_{|{\bf p-q}|}
\biggl[ 2\sin^2({\bf p\times q}l^2/2)
\Bigl[L_2(|{\bf p-q}|^2l^2/2)
+\bigl[L_1(|{\bf p-q}|^2l^2/2)\bigr]^2\Bigr]
\nonumber\\
&&
\mbox{\hspace{1in}}
+\Bigl[1-L_1(|{\bf p-q}|^2l^2/2)\Bigr]
\Bigl[L_1(|{\bf p-q}|^2l^2/2)-L_2(|{\bf p-q}|^2l^2/2)\Bigr]
\biggr],
\nonumber\\
&&
\tilde{V}_1^{01}({\bf p},{\bf q})=
2\cos({\bf p\times q}l^2)
e^{-|{\bf p-q}|^2l^2/2}v_{|{\bf p-q}|}
\frac{|{\bf p-q}|^2l^2}{4}L_1^1(|{\bf p-q}|^2l^2/2),
%\nonumber
\end{eqnarray}
with
% \begin{eqnarray}
% %\label{eq-n=1}
% &&
% L_1(|{\bf p-q}|^2l^2/2)=1-|{\bf p-q}|^2l^2/2,
% %\nonumber\\
% %&&
% ~~~
% L_1^1(|{\bf p-q}|^2l^2/2)=2-|{\bf p-q}|^2l^2/2,
% \nonumber\\
% &&
% L_2(|{\bf p-q}|^2l^2/2)=1-|{\bf p-q}|^2l^2+|{\bf p-q}|^4l^4/8,
% \nonumber
% \end{eqnarray}
% and 
$v_q=2\pi e^2/q$, while the single-exciton energies 
$E_{1p}^{\alpha \beta} \equiv E_{1p}^{\alpha \beta\alpha \beta}$ have the following form
\begin{eqnarray}
\label{en-n=1}
&&
E_{1p}^{00}=-\sqrt{\frac{\pi}{2}}\frac{e^2}{l}
\Biggl[\Phi\Bigl(\frac{1}{2},1,-\frac{p^2l^2}{2}\Bigr)
-\Phi\Bigl(\frac{3}{2},1,-\frac{p^2l^2}{2}\Bigr)
+\frac{3}{4}\Phi\Bigl(\frac{5}{2},1,-\frac{p^2l^2}{2}\Bigr)\Biggr],
\nonumber\\
&&
E_{1p}^{-1-1}=-\sqrt{\frac{\pi}{2}}\frac{e^2}{l}
\Phi\Bigl(\frac{1}{2},1,-\frac{p^2l^2}{2}\Bigr),
\nonumber\\
&&
E_{1p}^{11}=-\sqrt{\frac{\pi}{2}}\frac{e^2}{l}
\Biggl[\Phi\Bigl(\frac{1}{2},1,-\frac{p^2l^2}{2}\Bigr)
-2\Phi\Bigl(\frac{3}{2},1,-\frac{p^2l^2}{2}\Bigr)
+\frac{15}{4}\Phi\Bigl(\frac{5}{2},1,-\frac{p^2l^2}{2}\Bigr)
\nonumber\\
&&
\mbox{\hspace{2in}}
-\frac{15}{4}\Phi\Bigl(\frac{7}{2},1,-\frac{p^2l^2}{2}\Bigr)
+\frac{105}{64}\Phi\Bigl(\frac{9}{2},1,-\frac{p^2l^2}{2}\Bigr)
\Biggr],
\nonumber\\
&&
E_{1p}^{01}=-\sqrt{\frac{\pi}{2}}\frac{e^2}{l}
\Biggl[\Phi\Bigl(\frac{1}{2},1,-\frac{p^2l^2}{2}\Bigr)
-\frac{3}{2}\Phi\Bigl(\frac{3}{2},1,-\frac{p^2l^2}{2}\Bigr)
+\frac{15}{8}\Phi\Bigl(\frac{5}{2},1,-\frac{p^2l^2}{2}\Bigr)
\nonumber\\
&&
\mbox{\hspace{3.5in}}
-\frac{15}{16}\Phi\Bigl(\frac{7}{2},1,-\frac{p^2l^2}{2}\Bigr)
\Biggr],
\nonumber\\
&&
E_{1p}^{0-1}=-\sqrt{\frac{\pi}{2}}\frac{e^2}{l}
\Biggl[\Phi\Bigl(\frac{1}{2},1,-\frac{p^2l^2}{2}\Bigr)
-\frac{1}{2}\Phi\Bigl(\frac{3}{2},1,-\frac{p^2l^2}{2}\Bigr)
\Biggr],
\nonumber\\
&&
E_{1p}^{1-1}=-\sqrt{\frac{\pi}{2}}\frac{e^2}{l}
\Biggl[\Phi\Bigl(\frac{1}{2},1,-\frac{p^2l^2}{2}\Bigr)
-\Phi\Bigl(\frac{3}{2},1,-\frac{p^2l^2}{2}\Bigr)
+\frac{3}{8}\Phi\Bigl(\frac{5}{2},1,-\frac{p^2l^2}{2}\Bigr)\Biggr],
\nonumber\\
&&
\tilde{E}_{1p}^{01}=-\sqrt{\frac{\pi}{2}}\frac{e^2}{l}
\Biggl[2\Phi\Bigl(\frac{3}{2},1,-\frac{p^2l^2}{2}\Bigr)
-\frac{3}{2}\Phi\Bigl(\frac{5}{2},1,-\frac{p^2l^2}{2}\Bigr)
\Biggr],
\nonumber
\end{eqnarray}
$\Phi(a,b,z)$ being the confluent hypergeometric function,
%% \begin{eqnarray}
%% %\label{eq-n=1}
%% &&
%% \Phi\Bigl(\frac{1}{2},1,-\frac{p^2l^2}{2}\Bigr)
%% =\sum_{k=0}^{\infty}(-1)^k\frac{(2k-1)!!}{2^{2k}}\frac{(p^2l^2)^k}{(k!)^2},
%% \nonumber\\
%% &&
%% \Phi\Bigl(\frac{3}{2},1,-\frac{p^2l^2}{2}\Bigr)
%% =\sum_{k=0}^{\infty}(-1)^k\frac{(2k+1)!!}{2^{2k}}\frac{(p^2l^2)^k}{(k!)^2},
%% \nonumber\\
%% &&
%% \Phi\Bigl(\frac{5}{2},1,-\frac{p^2l^2}{2}\Bigr)
%% =\frac{1}{3}\sum_{k=0}^{\infty}(-1)^k\frac{(2k+3)!!}{2^{2k}}
%% \frac{(p^2l^2)^k}{(k!)^2},
%% \nonumber\\
%% &&
%% \Phi\Bigl(\frac{7}{2},1,-\frac{p^2l^2}{2}\Bigr)
%% =\frac{1}{15}\sum_{k=0}^{\infty}(-1)^k\frac{(2k+5)!!}{2^{2k}}
%% \frac{(p^2l^2)^k}{(k!)^2},
%% \nonumber\\
%% &&
%% \Phi\Bigl(\frac{9}{2},1,-\frac{p^2l^2}{2}\Bigr)
%% =\frac{1}{105}\sum_{k=0}^{\infty}(-1)^k\frac{(2k+7)!!}{2^{2k}}
%% \frac{(p^2l^2)^k}{(k!)^2},
%% \nonumber
%% \end{eqnarray}
%% with $(-1)!!=1$. 
and the pump-induced polarization is given by
\begin{eqnarray}
P_1(t)=-i\mu\int_{\infty}^{t}dt'e^{-i(\Omega_1+E_{10}^{00}-i\Gamma)(t-t')}
{\cal E}_2(t').
%\nonumber
\end{eqnarray}
The first equation in systems (\ref{eq-n=1}) describes the time-evolution of
two {\em e-h} pairs, excited to $n=1$ LL by the pump pulse
[Fig. \ref{fig1}(a)], with and without
inter-LL scattering (second and third terms, respectively). The third and
second equations in  (\ref{eq-n=1}) describe the similar time evolution of the
state (c) in Fig. \ref{fig1} and of the state obtained from (c) by exchanging
the holes in the valence band. The last terms of the first three
equations in (\ref{eq-n=1}) describe the Auger scattering that relates the
corresponding states to the state (b) in  Fig. \ref{fig1}, described by the
fourth equation. 
Note that only state (b) lacks the {\em e-h} symmetry, i. e., does not
transform into itself under replacement of electrons by hole and vice-versa in
Fig. \ref{fig1}, as indicated by the source term in the rhs of the forth
equation in (\ref{eq-n=1}). Its amplitude, $w_{1q}^{01}(t)$, plays the role of
the source for the rest of the system (\ref{eq-n=1}) (last terms) and thus
represents the sole source for the interaction-induced polarization.

In the numerical calculations below, we consider resonant excitation, i.e.,
central frequency tuned at the transitions between ground state and $n=1$
MX with binding energy 
\begin{eqnarray}
E_{10}^{00}=-\frac{3}{4}E_0,
~~~
E_0=\sqrt{\frac{\pi}{2}}\frac{e^2}{l},
%\nonumber
\end{eqnarray}
where $E_0$ is the binding energy of the $n=0$ MX.

% Note that 
% \begin{equation}
% E_{10}^{-1-1}=-E_0,
% ~~~
% E_{10}^{00}=-\frac{3E_0}{4},
% ~~~
% E_{10}^{11}=-\frac{41E_0}{64},
% \end{equation}
% so that $E_{10}^{11}+E_{10}^{11}-2E_{10}^{00}=-9E_0/64$.

%%%%%%%%%%%%%%%%%%%%%%%%%%%%%%%%%%%%%%%%%%%%%%%%%

The above equations for $w$ have been solved numerically using fourth-order
Runge-Kutte routine. FWM polarization was then calculated from
Eqs. (\ref{FWM-gen}-\ref{PB}) with $n=1$, for Gaussian pump and probe pulses
with the duration $t_p$ separated by time delay $\tau$. Dephasing due to
electron-phonon interactions have been accounted for by MX width
$\Gamma$ and by the width $\gamma$ characterizing the dephasing of two-pair
system. We emphasize that in the case of interacting excitons, $\gamma$ does
not necessarily equal $2\Gamma$, as has been pointed out in experiment
\cite{kner98}. 

\begin{figure}[thb]
\centering
\includegraphics[width=4in]{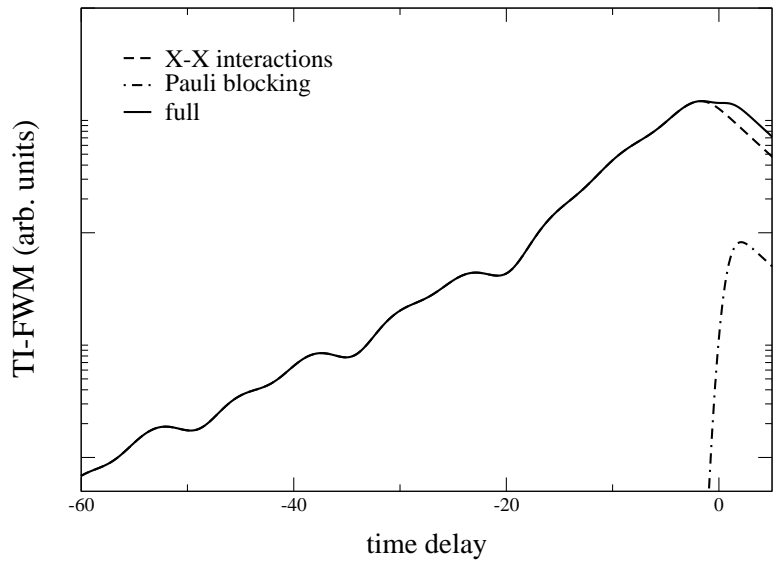}
\caption{\label{fig2}
Calculated TI-FWM signal versus time delay (in units of
  magnetoexciton energy) shows oscillations corresponding to the
  interference between states (a) and (b), and between states (a) 
  and (c) in Fig. \protect\ref{fig1}.
 }
\end{figure}

In Fig. \ref{fig2} we show calculated time-integrated FWM signal,
\begin{eqnarray}
\label{tifwm}
{\rm TI-FWM}=\int dt |{\cal P}(t)|^2,
%\nonumber
\end{eqnarray}
for parameter values  $t_p=\hbar/ E_0$, $\Gamma =0.1  E_0$,  $\gamma =0.05  E_0$
versus time delay (in units of $E_0$). The dot-dashed curve corresponds to
Pauli blocking contribution that comes from non-interacting excitons. It
should be emphasized that, in strong magnetic field, the {\em only}
interaction-induced contribution comes from Auger scattering, so the
Hartree-Fock contribution is suppressed by a small parameter
$E_0/\hbar\omega_c\ll 1$. 
It can be seen that Auger scattering of magnetoexcitons strongly enhances the
amplitude of FWM signal. For negative time delays, TI-FWM signal develops an
exponential (with decay-time $\hbar/\gamma$) tail that is characteristic
for interacting excitons. Furthermore, TI-FWM signal exhibits two sets
of oscillations superimposed on each other. These oscillations represent, in
fact,  quantum beats between four-particle configurations contributing to FWM
polarization. We identify more the pronounced oscillations with the
interference between states excited state (a) and state (b) in Fig. \ref{fig1}
related to (a) by electron Auger-scattering in conduction band. These states
are characterized by Coulomb energies [see Eqs. (\ref{eq-n=1}) and
(\ref{en-n=1})] 
$E_a=2E_{10}^{00}=-\frac{3}{2} E_0$ and
$E_b=E_{10}^{01}+E_{10}^{0-1}=-\frac{15}{16} E_0$, 
so the oscillations period corresponds to the energy difference
$E_b-E_a=\frac{9}{16}E_0$. 
The weaker oscillations originate from the interference between states
(a) and (c) in Fig. Fig. \ref{fig1}, where the state (c) is
characterized by energy $E_c=2E_{10}^{1-1}=-\frac{3}{4}E_0$, so the
period is determined by  $E_c-E_a=\frac{3}{4}E_0$. Note that the state (d),
obtained from (c) by hole exchange in the valence band, is characterized by
the energy  
$E_d=E_{10}^{11}+E_{10}^{-1-1}=-\frac{105}{64} E_0$ and the
corresponding period, determined by $E_d-E_a=\frac{9}{64}E_0$, is too
long to be noticeable in Fig. \ref{fig2}. It should be emphasized,
however, that the above estimates involve zero-momentum energies of
each of the pairs that constitute four-particle correlated state. In
fact, Auger processes are accompanied by a momentum exchange between
MX's , so only the total momentum of a four particle correlation is
zero. This momentum exchange leads to the damping of oscillations. 
Since (a) and (c) are related via two Auger processes (in conduction 
and in valence band) the additional momentum exchange during the hole 
Auger-scattering in valence band leads to a much much stronger
damping of (a)-(c) oscillations as compared to (a)-(b) oscillations.

\section{conclusions}
\label{conc}

We investigated the role  of inter-LL transitions in the coherent
dynamics of quantum well excitons in strong magnetic field. While 
on the lowest LL, the suppression of inter-LL transitions results
in the absence of exciton-exciton interactions, on higher LL levels
the interactions become dominant due to resonant Auger processes.
The coherent signature of exciton  Auger-scattering can be traced in
the FWM polarization as quantum beats corresponding to
the interference between optically-inactive  four-particle
correlated states.

%\acknowledgments
Author thanks I. E. Perakis for useful discussions and N. Primozich
for help in numerical calculations. 
This work was supported by National Science Foundation under
grants DMR-0304036 and DMR-0305557, and by Army Research Laboratory
under grant DAAD19-01-2-0014.

%% Army High Performance Computing
%% Research Center under the auspices of the Department of the Army,
%% Army Research Laboratory under Cooperative Agreement
%% No DAAD19-01-2-0014, 
%% the content of which does not necessarily
%% reflect the position or the policy of the government, and no official
%% endorsement should be inferred; 

% \begin{references}

%\end{references}

\end{document}